# REPRINT

## Chapter 16

# Multi-objective design of multilayer microwave dielectric filters using artificial bee colony algorithm

Abdurrahim Toktas

Department of Electrical and Electronics Engineering,
Faculty of Engineering, Karamanoglu Mehmetbey University,
Karaman, Turkey, atoktas@kmu.edu.tr

**Abstract.** Artificial bee colony algorithm (ABC) developed by inspiring the foraging phenomena of the natural honey bees is a simple and powerful metaheuristic optimization algorithm. The performance of single objective ABC performance has been well demonstrated by implemented to different design optimization problems from electrical and mechanical to civil engineering. Its efficacy is anticipated in the multi-objective design optimization of multilayer microwave dielectric filters (MMDFs) which is a challenging multi-objective problem falls into the context of electromagnetic (EM) design in electrical engineering. The MMDF is composed of superimposed multiple dielectric layers in order to pass or stop the EM wave at the desired frequency bands. An accurate dual-objective EM model of MMDF is constituted to be the total reflection coefficients for oblique incident wave angular range with transverse electric (TE) or transverse magnetic (TM) polarizations. The objective functions depend on the total reflection (TR) at the pass and stop band regions. Three types of the MDFs that are low pass (LP), high pass (HP) and band pass (BP) MMDFs are optimally designed across the constituted dual-objective EM model with a frequency dependent material database through using the multi-objective ABC (MO-ABC) strategy. The Pareto optimal solutions are refined within the possible solutions for synchronously minimizing the objective functions. The global optimal MMDF designs are selected from the Pareto optimal solutions by providing the trade-off therein between the objective functions. The performance study regarding the frequency characteristics of the designed MMDFs are well presented via numerical and graphical results. It is hence demonstrated that MO-ABC is even versatile and robust in the multi-objective design of MMDFs.

## 16.1 Introduction

Metaheuristic optimization algorithms which have been developed by inspiring natural phenomenon have achieved amazing results in optimum engineering designs in the last decades. They derive their achievements from the perfection in nature. Therefore, the more similar the algorithm is modeled to the processing of the nature, the more successful it is. Almost all-natural phenomenon has attempted to model for an optimization algorithm [1]. In general, the algorithms are built on controllable stochastic computations that iteratively try to improve the candidate solutions. The most prominent are herein: Genetic algorithm (GA) [2] and differential evolution (DE) [3] were developed by inspiring the mutation phenomenon of living organism. Particle swarm optimization (PSO) was simulated the flock behavior organisms such as birds and fishes [4]. Ant colony optimization (ACO) was mimicked the communication and direction finding paradigm of the biological ants [5]. Harmony search (HS) was inspired by the improvisation process of jazz musicians [6]. Cuckoo search (CS) was modelled the obligate brood parasitism of some cuckoo species by laying their eggs in the nests of other species [7]. Bat algorithm (BA) was simulated the echolocation behavior of the natural bats [8]. Artificial bee colony (ABC) was emerged by making analogy of nectar foraging behavior of the honey bees [9]. ABC algorithm and its variants has been successfully applied to a variety of the engineering problems [10–14]. Although several multi-objective versions of ABC have been proposed [11, 12, 15–17]¶, its success on multi-objective electromagnetic (EM) computational expensive problems has been still remaining a curiosity. When metaheuristic optimization algorithms are first developed, they are emerged for single objective optimizations. Additional techniques, i.e. Pareto optimality are integrated them to adapt multi-objective capability, in general [18–20]. Thanks to Pareto optimality, it is possible to refine optimal solutions according to all objective vectors and herewith to determine a global optimal solution by taking account of the trade-offs among the objective vectors.

Microwave dielectric filters (MDFs) are interested components for microwave and wireless applications such as communication, antenna, radar and biomedical systems [21]. MDFs are used to reflect, transmit or absorb EM fields at the desired frequency band of the field. The most commonly designed MDFs are low pass (LP), high pass (HP) and band pass (BP) filters. It is desired that the incident EM wave passes through the layers of the MDF at the pass band and reflects from the MDF at the stop band [22]. It is inconvenient to obtain near ideal filter characteristic using single layer MDF design. On the other hand, by using multilayer MDF (MMDF) design composed of different dielectric layers, it is more likely to achieve rigorous and sharp filter characteristic. The design of MMDF requires a multi-objective optimization strategy in which the reflected EM wave must be synchronously minimized at the pass band and maximized at the stop band. A computational EM model is also required for this multi-objective optimization scheme, taking into account the incident wave angle and polarization at all layer interfaces. The design parameters of the MMDF regarding both the thickness and material types of the layers should be simultaneously determined by searching within an existing material database with frequency dependent complex permittivity through a powerful multi-objective optimization procedure as well as considering the trade-off between the objective functions.

Few studies have attempted to design MMDFs in recent years. In [23], a procedure for designing MMDFs was presented a binary-coded GA. Several five-layer MMDFs were designed between 8 GHz and 18 GHz for single objective through non-defined dielectric materials. In other words, dielectric constants for layer materials were searched continuously between 1.03 and 10.0. The two objectives were combined by weighting each of them. In [24], an evolutionary programming algorithm was used to design MMDFs between 24 GHz and 36 GHz for single objective as similar to the procedure in [23]. Dielectric constants were Rogers Corporation with values ranging from 1.01 to 10.2. In [25], a few MMDFs were designed using multi-objective GA between 24 GHz and 36 GHz by selecting material types from a material database including dielectric constants. They are all designed at incident angle 45 degree. In real applications, material types are usually with frequency dependent complex permittivity and permeability. However, material database with only real permittivity (dielectric constant) was employed.

In this chapter, three types of five-layer MMDFs that have LP, HP and BP filter characteristics are designed through multi-objective ABC (MO-ABC) algorithm. Pareto optimality that allows to refine optimal solution in accordance with the multiple objective values is incorporated to ABC so as to adapt multi-objective ability. The materials are selected from a frequency dependent artificial material database including complex permittivity and permeability. The MMDF are designed to operate at the frequency range of 2-18 GHz. The LP filter passes the band of 2-10 GHz and stops the band of 10-18 GHz and the HP filter stops the band of 2-10 GHz and passes the band of 10-18 GHz. The BP filter stops both bands of 2-8 GHz and 12-18 GHz, and passes the band of 8-12 GHz. Two objective functions are constituted to be minimized the total reflections (TRs) at the pass and stop bands. Global optimal MMDFs are selected within the Pareto optimal solutions by ensuring the trade-off between the two objective functions. Eventually, the frequency characteristics of the designed MMDFs are elaborately examined to demonstrate the performance of MO-ABC.

## 16.2 MO-ABC algorithm

ABC modelled by observing the collective foraging behavior of the natural honey bees was initially originated for single objective optimization problems. Pareto optimality is a simple and easy approach to adapt multi-objective strategy to ABC. Pareto optimality obtains the optimal solutions which are the fittest ones according to either objective with respect to all objectives among the possible solutions.

### 16.2.1 Pareto optimality algorithm

Engineering design problems are often burdensome multi-objective optimizations. Since objective functions are dependent on decision vectors (design variables), they are hence related to each other, as well. In other words, while one objective is improved by changing the decision vector, the others may deteriorate [26]. Therefore, the results of all objective functions must be taken into consideration in finding out optimal solutions with respect to all objectives. Given that every decision vector has an outcome in terms of each objective function. A decision vector may dominate the other vectors in accordance with one objective function value, whereas it may not dominate according with the other objective function values. Hence a set of decision vectors that dominate all other vectors for entire objective space should be considered. Pareto optimality whose pseudocode is given in Algorithm 1 is the most effective way that is incorporated to the metaheuristic algorithms, allowing to independently determine a diverse and uniform optimal solution set (non-dominated solutions) so-called Pareto front. Assuming that there is objective space $OF(\mathrm{x}) = [of_1, of_2, \ldots, of_N]$ with $N$ ($k$=1, 2, ..., $N$: $N \geq 2$) objective functions $of_k(x)$. The decision vector is $x_j = [x_1, x_2, \ldots, x_d]$ where $j$=1, 2, ..., $d$ called decision space and search space $x_i = [x_1, x_2, \ldots, x_{NP}]$ with $i$=1, 2, ..., $NP$ where $NP$ is number of population.

**Algorithm 1** Pseudocode of Pareto optimality

```
x* and x# are decision vectors
If x* dominates x
    x# = x*
    i = 0
else if x# dominates x*
    i = i +1
else if x* and x# are non–dominated (Pareto optimal) solutions
    if rand < 0.5
    x# = x*
    i = 0
    end if
end if
```

### 16.2.2 ABC algorithm

ABC algorithm was developed by simulating the self-organizing and collective foraging intelligence of the honey bee colonies [9, 27]. The bees forage as three colonies: employed bee, onlooker bee and scout bee. The phases of ABC are also referred to the same name of colonies. In colonies, artificial bees search for quality nectar sources. ABC whose pseudocode given in Algorithm 2 proceeds iteratively through these phases. The total number of bees in the colony whose number is the $NP$ is considered as two equal colonies: the employed and onlooker bees. Each bee works for a nectar source of which location stands for a candidate solution. The qualities of nectar sources correspond to the fitness of the candidate solutions. Number of $NP/2$ candidate solutions are evaluated at both employed and onlooker bee phases, separately. All employed bees are initially appointed as scout bees to discover new nectar sources. Then the employed bees search around the nectar sources and transfer information regarding the qualities of nectar sources to the onlooker bees. According to this information, the onlooker bees search again in vicinities of the nectar sources determined by the employed bees. If a more quality nectar source could not be found around the nectar source after a specified number of trying

"*limit*", it then abandoned and a totally new nectar source is defined instead of the old one.

**Algorithm 2** Pseudocode of MO-ABC

Begin<br>
**Initialization**<br>
Set control parameters: limit, *NP* and *NI*<br>
Generate randomly initial populations (decision vectors) with *NP* (16.1)<br>
Compute initial objective vectors and fitness vectors (16.2)<br>
**Employed bee phase**<br>
Change randomly selected decision vectors (candidate solutions) (16.3)<br>
Compute the new objective vectors and fitness vectors (16.2)<br>
Compute the probability vectors (16.4)<br>
*Refine Pareto optimal solutions (Algorithm 1)*<br>
**Onlooker bee phase**<br>
Change randomly selected candidate solutions (16.3)<br>
Compute the new objective vectors and fitness vectors using (16.2)<br>
*Refine Pareto optimal solutions (Algorithm 1)*<br>
  If a solution cannot be improved after trying as number of *"limit"*<br>
  **Scout bee phase**<br>
  Replace the old solution with a new generated solution (16.1)<br>
  Compute the new objective vectors and fitness vectors (16.2)<br>
  Record the best solution obtained so far<br>
End

In ABC algorithm, at the initialization, candidate solution vectors (decision vectors) with *NP* $x_{ij}$ are randomly defined within the decision space constrained between $x_j^{\max}$ and $x_j^{\min}$ by using the following mathematical operator.

$$x_{ij} = x_j^{\min} + rand(0,1)(x_j^{\max} - x_j^{\min}) \tag{16.1}$$

The objective vectors $of_i(x)$ and accordingly the fitness vectors $fit_i$ assessing the quality of candidate solutions are determined for all objective functions to evaluate the quality of candidate solutions as follows

$$fit_i = \begin{cases} \dfrac{1}{1+of_i(x)} & if\ of_i(x) \geq 0 \\ 1 + abs\left[of_i(x)\right] & if\ of_i(x) < 0 \end{cases} \tag{16.2}$$

In employed bee phase, new solution vector $v_{ij}$ is produced near the before founded candidate solutions $x_{ij}$ and $x_{kj}$, which are randomly selected from the population using the following operator.

$$v_{ij} = x_{ij} + \phi_{ij}(x_{ij} - x_{kj}) \tag{16.3}$$

The objective vectors $of_i(v)$ and fitness vectors $fit_i$ are repeatedly evaluated for the new decision vector. The Pareto optimal solutions are obtained using Pareto optimality (Algorithm 1). A probability vectors $prob_i$ through the fitness vectors are evaluated for all objectives via the operator (16.4). The Pareto optimal solutions are then obtained depending on Pareto optimality (Algorithm 1).

$$prob_i = \frac{fit_i}{\sum_{n=1}^{NP/2} fit_n} \tag{16.4}$$

In onlooker bee phase, new solutions are regenerated in the vicinity of the before obtained solutions $x_{ij}$ and $x_{kj}$ using the operator in the employed bee phase (16.3). The objective vectors $of_i(v)$ and fitness vectors $fit_i$ are then calculated via the operator (16.2) with similar manner in the Employed bee phase. New Pareto optimal solutions are hence refined with regard to the new objective vector $v_{ij}$ by repeating the processes in Algorithm 1.

In scout bee phase, a totally new candidate solution is stochastically regenerated via the same operator (16.1) in place of a solution that could not be improved after the number of trying *"limit"*. The objective vectors $of_i(v)$ and fitness vectors $fit_i$ are therefore calculated via the operator (16.2). Finally, the last Pareto optimal solutions achieved so far are saved. Those three phases are iteratively proceeded up to a predefined number of iterations (NI).

## 16.3 Multi-objective EM model of the MMDF

The MMDF is composed of multiple dielectric layers that stacked on each other. A MMDF structure with *n* layers $i = 0,1,2,...,n$ is illustrated in Fig. 16.1. Each layer has thickness of $d_i$ and material of $m_i$ having complex permittivity $\varepsilon_i$ and permeability $\mu_i$. The propagating wave through the air strikes onto Interface 1 with the incident angle of $\theta$. The relationship between the incidence angle $\theta_{i-1}$ and transmitted incident waves angle $\theta_i$ at each interface can be determined by Snell's law as follows

$$\frac{\sin\theta_i}{\sin\theta_{i-1}} = \sqrt{\frac{\mu_{i-1}\varepsilon_{i-1}}{\mu_i\varepsilon_i}} \tag{16.5}$$

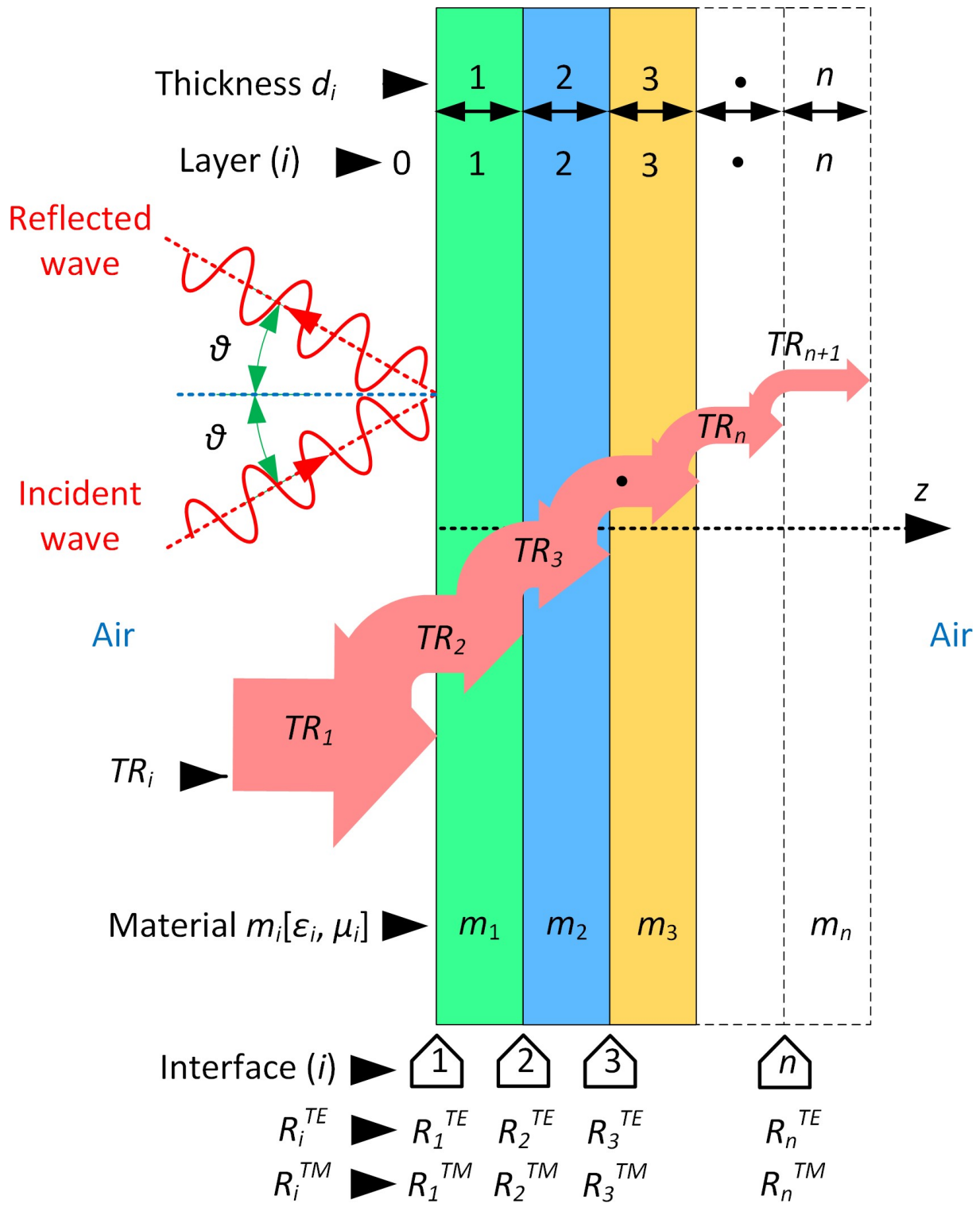

**Fig. 16.1.** The structure of conceptual MMDF

An EM model regarding the for total reflection (TR) coefficient taking account of the oblique incident wave angle and comprising sub-reflections at the inner interfaces together with transverse electric (TE) and transverse (TM) at each layer is given below [15].

$$TR_i^{TE} = \frac{R_i^{TE} + \left(TR_{i+1}^{TE}\right)e^{-2jkz_i d_i}}{1 + R_i^{TE}\left(TR_{i+1}^{TE}\right)e^{-2jkz_i d_i}} \tag{16.6}$$

$$TR_i^{TM} = \frac{R_i^{TM} + \left(TR_{i+1}^{TM}\right)e^{-2jkz_i d_i}}{1 + R_i^{TM}\left(TR_{i+1}^{TM}\right)e^{-2jkz_i d_i}} \tag{16.7}$$

where, $kz_i = \cos\theta_i \omega \sqrt{\mu_i \varepsilon_i}$ is the complex wave propagation number along the *z*-direction and $\omega = 2\pi f$ is the angular frequency, *f* is the frequency of the incident wave. The sole interface reflection (*R*) coefficients at the inner interfaces are given as

$$R_i^{TE} = \frac{\mu_i kz_{i-1} - \mu_{i-1} kz_i}{\mu_i kz_{i-1} + \mu_{i-1} kz_i} \tag{16.8}$$

$$R_i^{TE} = \frac{\varepsilon_i kz_{i-1} - \varepsilon_{i-1} kz_i}{\varepsilon_i kz_{i-1} + \varepsilon_{i-1} kz_i} \tag{16.9}$$

### 16.3.1 The dual objective functions for the design of MMDFs

The three types of MMDF having LP, HP and BP characteristics are designed through MO-ABC algorithm. The MMDFs is aimed to operate between 2 GHz and 18 GHz. The band characteristics of the considered MMDFs are depicted in Fig. 16.2 [28]. In this regard, the objective functions of LP filter with cut-off frequency 10 GHz are constituted so as to pass the band of 2-10 GHz and to stop the band of 10-18 GHz. Conversely, those of HP filter are formed so that it passes the band of 10-18 GHz and stopes the band of 2-10 GHz. Eventually, the BP filter with cut-off frequencies 8 GHz and 12 GHz is constructed to pass the band of 8-12 GHz and to stop the bands of 2-8 GHz and 12-18 GHz. The TR at 0.316= –10 dB will be considered for a reference level of band pass region, meaning that 90% of the incident wave power pass through the MMDF.

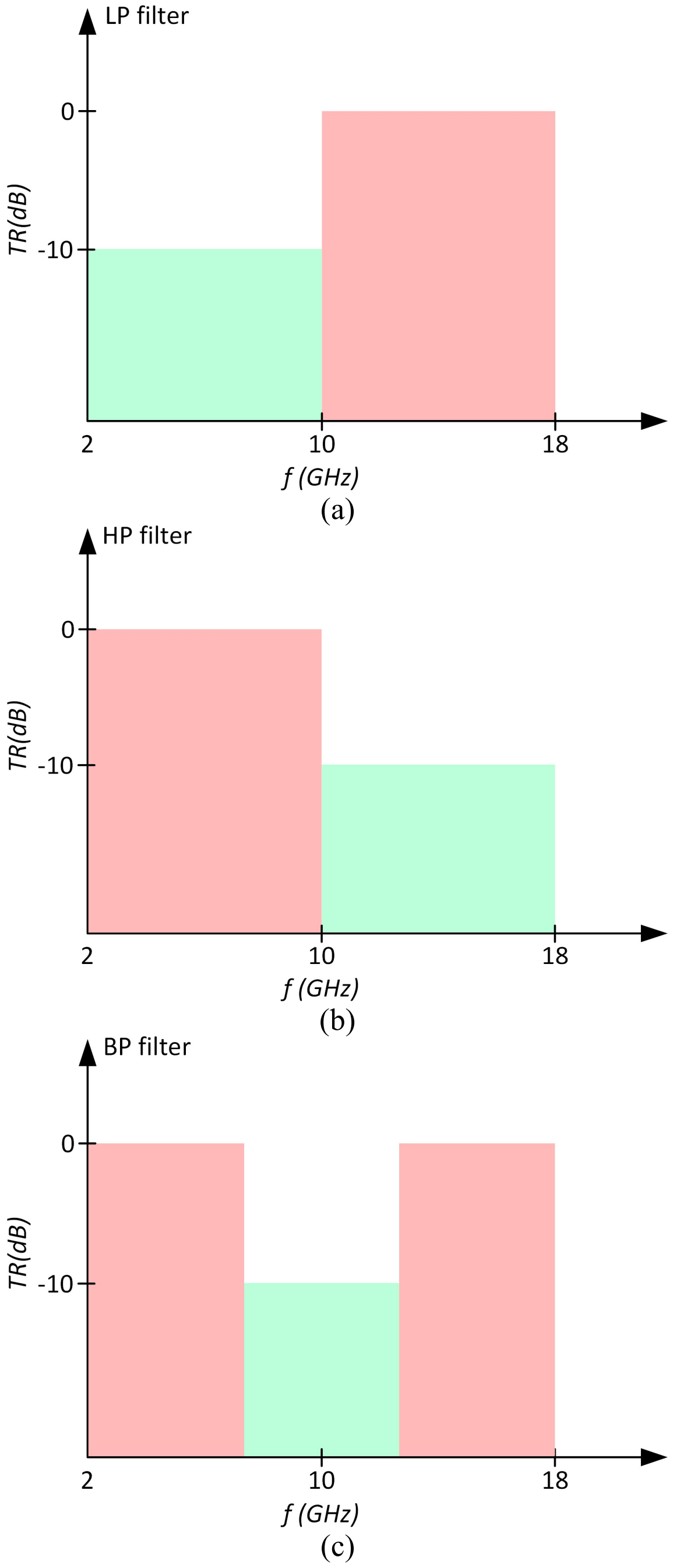

**Fig. 16.2.** The frequency band characteristics of the considered MMDFs: a) LP filter, b) HP filter, c) BP filter

The two objective functions regarding the mean of TR at the pass band and stop band are separately constituted to be minimizing. At pass band it is aimed to minimize the first objective function $of_1$, however it is desired to maximize the second objective function $of_2$. Since the numerical value of TR varies between 0 and 1, the second objective function can be converted to be minimized by subtracting each $TR^{TE}$ and $TR^{TM}$ from 1. The final objective functions that should be minimized are constituted as follows

$$of_1 = \frac{\sum_{f_l}^{f_u} \frac{\sum_{\theta_l}^{\theta_u}\left[\left|TR_1^{TE}(\theta, f_p)\right| + \left|TR_1^{TM}(\theta, f_p)\right|\right]}{2N_{ap}}}{N_{fp}} \tag{16.10}$$

$$of_2 = \frac{\sum_{f_l}^{f_u} \frac{\sum_{\theta_l}^{\theta_u}\left[2 - \left|TR_1^{TE}(\theta, f_s)\right| - \left|TR_1^{TM}(\theta, f_s)\right|\right]}{2N_{ap}}}{N_{fp}} \tag{16.11}$$

where, $N_{ap}$ is the number of angle points for the range of $\theta_l \le \theta \le \theta_u$ and $N_{pf}$ is the number of frequency points for the range of pass or stop bands. Therefore, the mean of the TR is obtained by divided them into $N_{ap}$ and $N_{pf}$. Moreover, because there are two TRs related to TE and TM, first they divided into 2. $f_p$ and $f_s$ are referred to the frequency point falls in pass and stop regions, respectively. Hence the range of $f_p$ and $f_s$ can be defined as follows

For the LP filter:

$$2\,GHz \le f_p \le 10\,GHz \tag{16.12}$$

$$10\,GHz \le f_s \le 18\,GHz \tag{16.13}$$

For the HP filter:

$$10\,GHz \le f_p \le 18\,GHz \tag{16.14}$$

$$2\,GHz \le f_s \le 10\,GHz \tag{16.15}$$

For the BP filter:

$$2\,GHz \le f_s \le 8\,GHz \quad (16.16)$$

$$8\,GHz \le f_p \le 12\,GHz \quad (16.17)$$

$$12\,GHz \le f_s \le 18\,GHz \quad (16.18)$$

## 16.4 The designed MMDFs through MO-ABC

Three types of five-layer MMDFs which are LP, HP and BP filters are designed in this study to operate between 2-18 GHz. It is aimed that the LP, HP and BP filters respectively pass the bands of 2-10 GHz, 10-18 GHz and 8-12 GHz, and stop the bands of 10-18 GHz, 2-10 GHz and 2-8/12-18 GHz. Through MO-ABC, the Pareto optimal solutions each of which stands for design parameters of the MMDFs are found out within the all possible solutions. The material types to be used in the dielectric layer are picked up from a predefined material database. Global optimal MMDFs are then selected among the Pareto optimal solutions.

### 16.4.1 The set parameters and material database

The range of the incident angle is between 0 and 45 degrees. The steps range for the angle and frequency are 15 degree and 0.2 GHz, respectively. Therefore, the $N_{ap}$ would be 4. $N_{fp}$ would be 41 for all bands of both designs of LP and HP filters and it would be 31 for both stop bands and 21 for the pass band of the BP filter. The thickness for each layer is searched between $d_l$ = 0 mm and $d_u$ = 3 mm. A frequency dependent material database which was initial defined in [29] are exploited in the multi-objective design of MMDFs. The database composed of 16 complex permittivity and permeability is tabulated in Table 16.1. On the other hand, the control parameters of ABC such as NP, NI and limit are set as 100, 1000 and 100, respectively.

**Table 16.1.** The predefined frequency dependent artificial material database

| Lossless dielectric materials ( $\mu' = 1$, $\mu'' = 0$ ) | | | | |
|---|---|---|---|---|
| Mat. # | $\varepsilon'$ | | | |
| 1 | 10 | | | |
| 2 | 50 | | | |
| Lossy magnetic materials ( $\varepsilon' = 15$ , $\varepsilon'' = 0$ ) $\mu'(f) = \frac{\mu'(1GHz)}{f^{\alpha}}$ ..... $\mu''(f) = \frac{\mu''(1GHz)}{f^{\beta}}$ | | | | |
| Mat. # | $\mu'(1GHz)$ | $\alpha$ | $\mu''(1GHz)$ | $\beta$ |
| 3 | 5 | 0.974 | 10 | 0.961 |
| 4 | 3 | 1.000 | 15 | 0.957 |
| 5 | 7 | 1.000 | 12 | 1.000 |
| Lossy dielectric materials ( $\mu' = 1$, $\mu'' = 0$ ) $\varepsilon'(f) = \frac{\varepsilon'(1GHz)}{f^{\alpha}}$ ..... $\varepsilon''(f) = \frac{\varepsilon''(1GHz)}{f^{\beta}}$ | | | | |
| Mat. # | $\varepsilon'(1GHz)$ | $\alpha$ | $\varepsilon''(1GHz)$ | $\beta$ |
| 6 | 5 | 0.861 | 8 | 0.569 |
| 7 | 8 | 0.778 | 10 | 0.682 |
| 8 | 10 | 0.778 | 6 | 0.861 |
| Relaxation-type magnetic materials ( $\varepsilon' = 15$ , $\varepsilon'' = 0$ ) $\mu'(f) = \frac{\mu_m f_m^2}{f^2 + f_m^2}$ ..... $\mu''(f) = \frac{\mu_m f_m f}{f^2 + f_m^2}$ $f$ and $f_m$ in GHz | | | | |
| Mat. # | $\mu_m$ | | $f_m$ | |
| 9 | 35 | | 0.8 | |
| 10 | 35 | | 0.5 | |
| 11 | 30 | | 1.0 | |
| 12 | 18 | | 0.5 | |
| 13 | 20 | | 1.5 | |
| 14 | 30 | | 2.5 | |
| 15 | 30 | | 2.0 | |
| 16 | 25 | | 3.5 | |

### 16.4.2 The performance results of the designed MMDFs

The objective functions involving $TR^{TE}$ and $TR^{TM}$ are computed at the incident angles of 0, 15, 30 and 45 degrees. The material types *m*[*ε*, *μ*] which would be in the thickness range of 0-3 mm are searched within the material database given in Table 16.1 In the design of five-layer MMDFs, the obtained Pareto optimal solutions for the LP, HP and BP filters are distributed over two-dimensional (2D) objective spaces in Fig. 16.3. The blue solid circle symbol indicates the Pareto optimal solutions and the red star symbol stands for the selected global optimal solution (MMDF design).

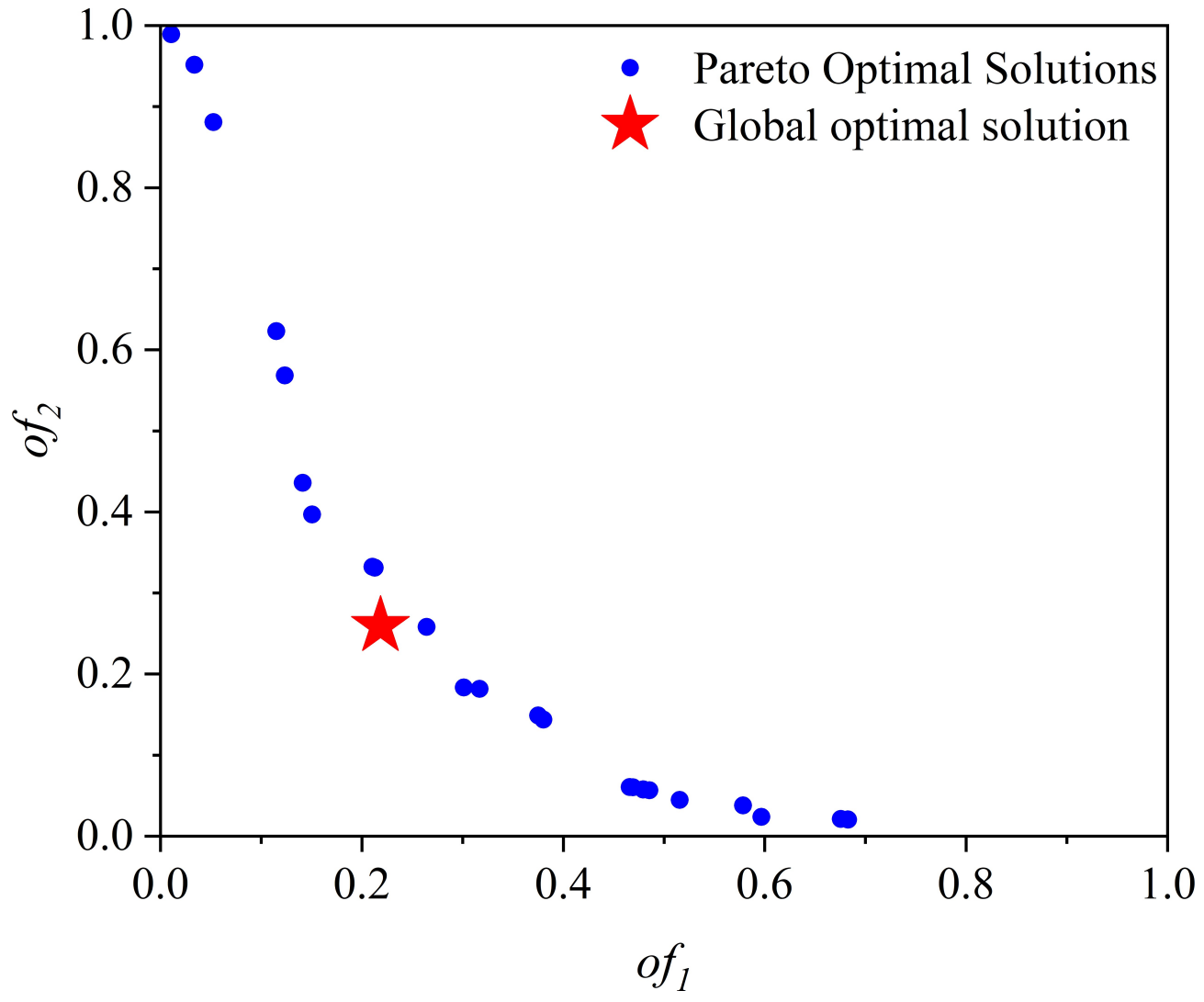

(a)

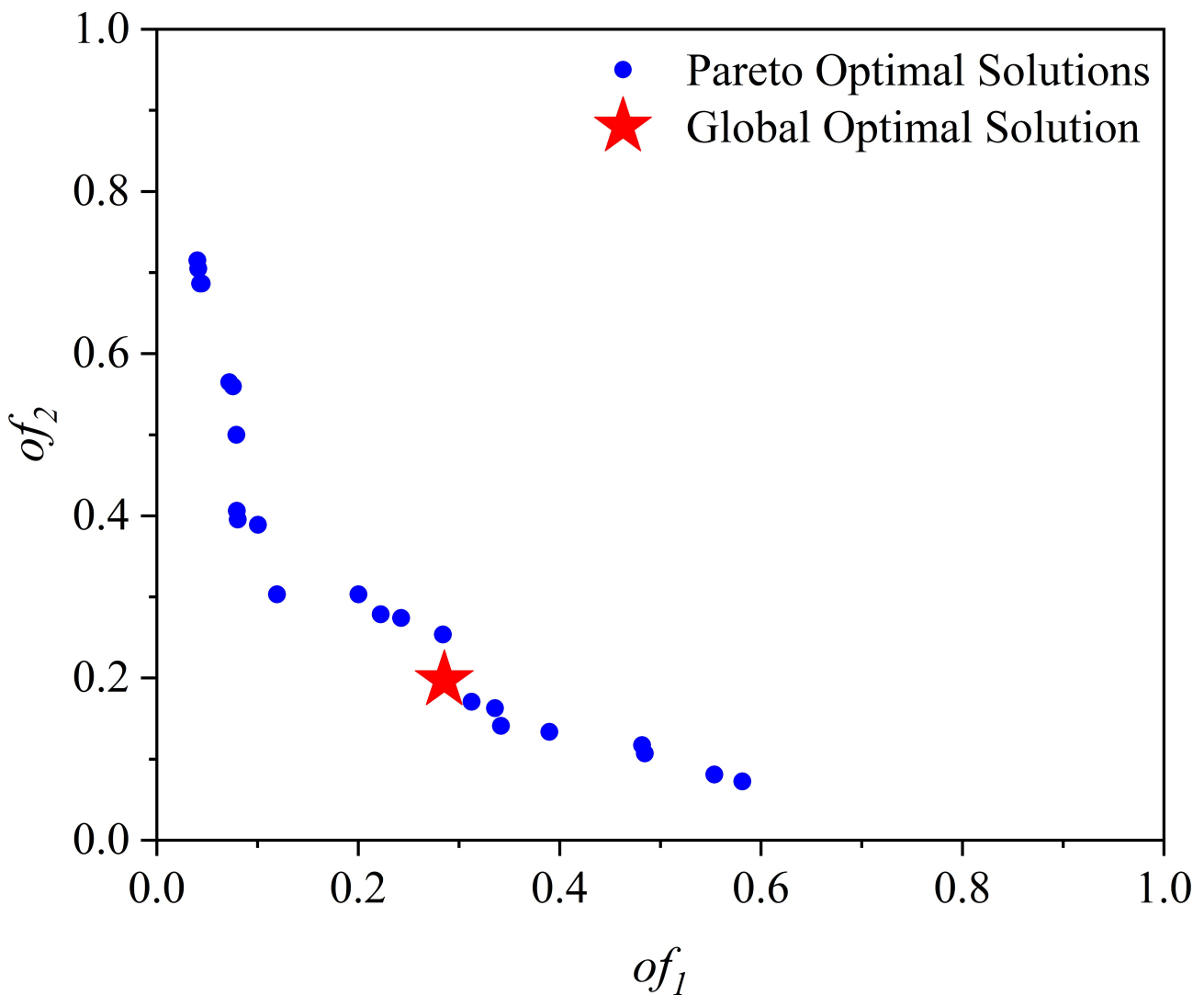

(b)

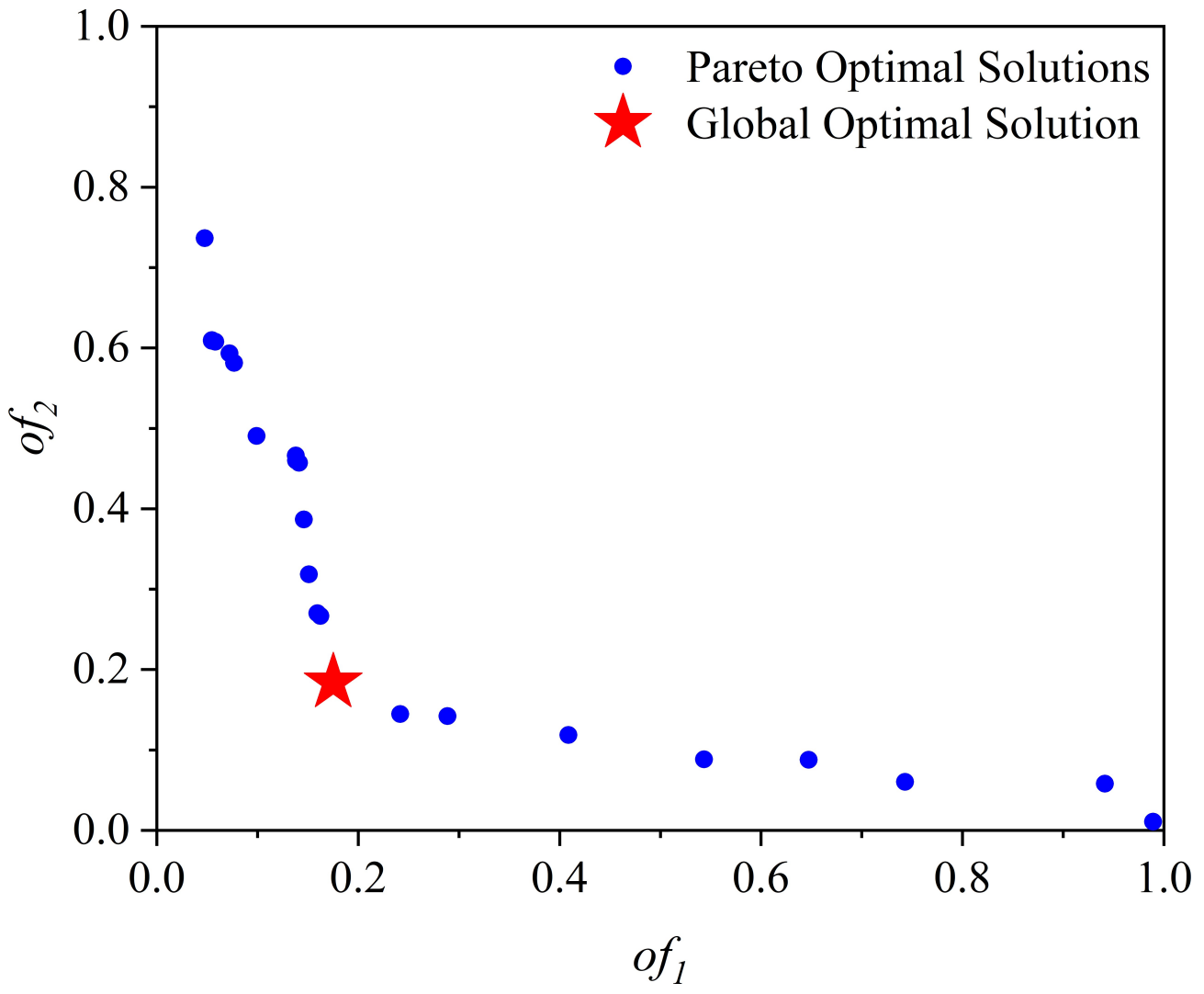

(c)

**Fig. 16.3.** 2D Pareto objective space for the design of MMDFs: a) LP filter, b) HP filter, c) BP filter

The global optimal MMDF the would be the trade-off solution with regard to both objective function values are selected to be diagonally the nearest to the origin of the 2D objective space. The selected global optimal

MMDFs with thickness, material types are tabulated in Table 16.2. The TT of the global optimal LP, HP and BP filters are 9.0799 mm, 5.5237 mm and 6.1891 mm, respectively. The TRs versus frequency characteristic for MMDFs at 0 (normal incidence) 15, 30 and 45 degrees are plotted in Fig. 16.4. It is seen that almost all TRs obey the desired MMDF characteristics. Yet the characteristics of HP filter is not coherent with an ideal filter as much as the other LP and HP filters. That is why the material database may not suitable for the HP filter. Only the $TR^{TE}$ at high degrees such as 30 and 45 are not under -10 dB. Because the $TR^{TE}$ is always higher the $TR^{TM}$.

**Table 16.2.** The selected global optimal MMDFs among the Pareto optimal solutions

| Layer sequence | LP filter | | HP filter | | BP filter | |
|---|---|---|---|---|---|---|
| | Global optimal solution | | Global optimal solution | | Global optimal solution | |
| | $of_1$ | $of_2$ | $of_1$ | $of_2$ | $of_1$ | $of_2$ |
| | 0.2185 | 0.2593 | 0.2855 | 0.1972 | 0.1751 | 0.1838 |
| | Mat. # | Thickness (mm) | Mat. # | Thickness (mm) | Mat. # | Thickness (mm) |
| 1 | 9 | 0.7118 | 1 | 0.2818 | 1 | 1.5615 |
| 2 | 8 | 3.0000 | 7 | 1.7758 | 6 | 0.3311 |
| 3 | 2 | 0.9224 | 8 | 0.2753 | 1 | 0.7746 |
| 4 | 8 | 3.0000 | 1 | 2.0800 | 2 | 0.9427 |
| 5 | 1 | 1.4457 | 13 | 1.1107 | 1 | 2.5793 |
| TT (mm) | 9.0799 | | 5.5237 | | 6.1891 | |

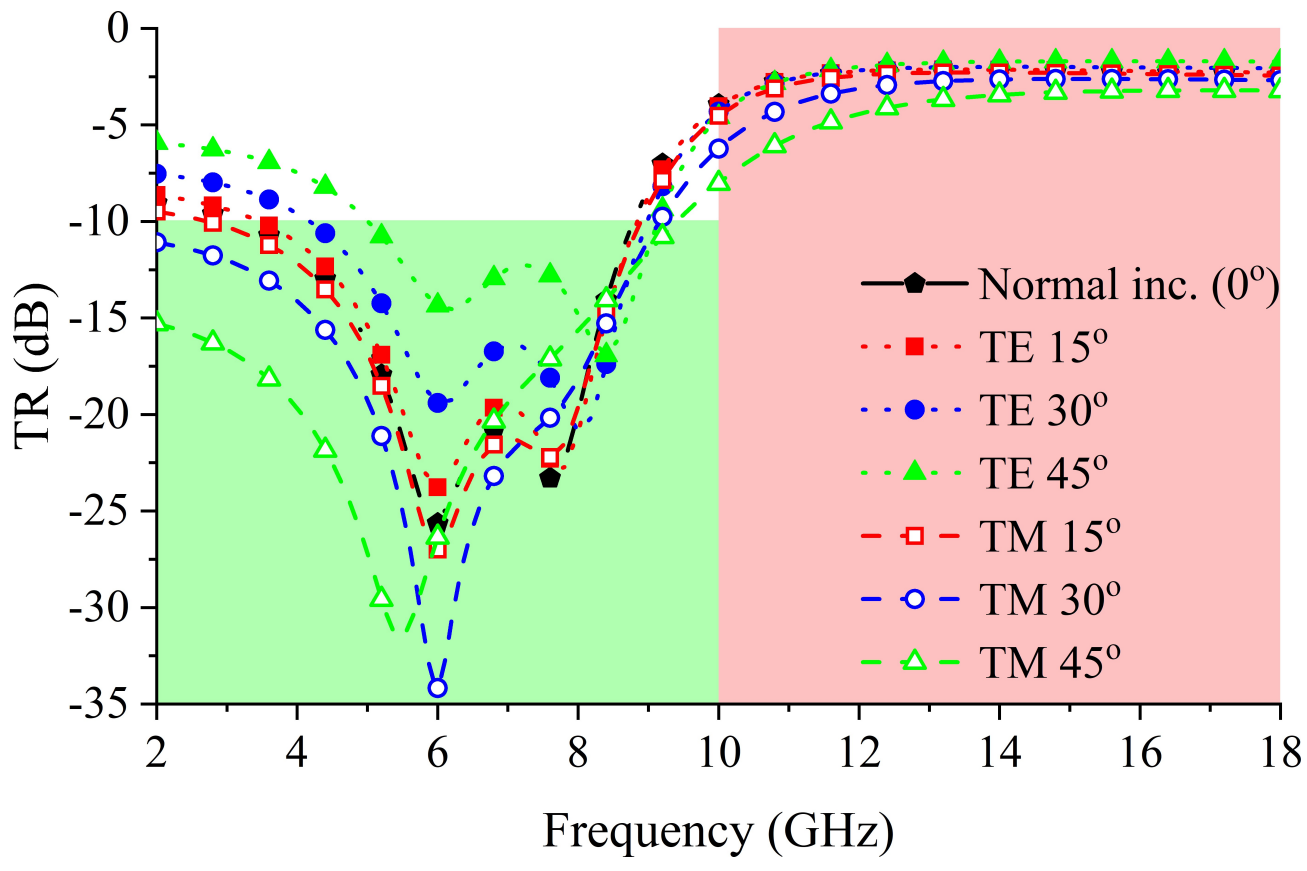

(a)

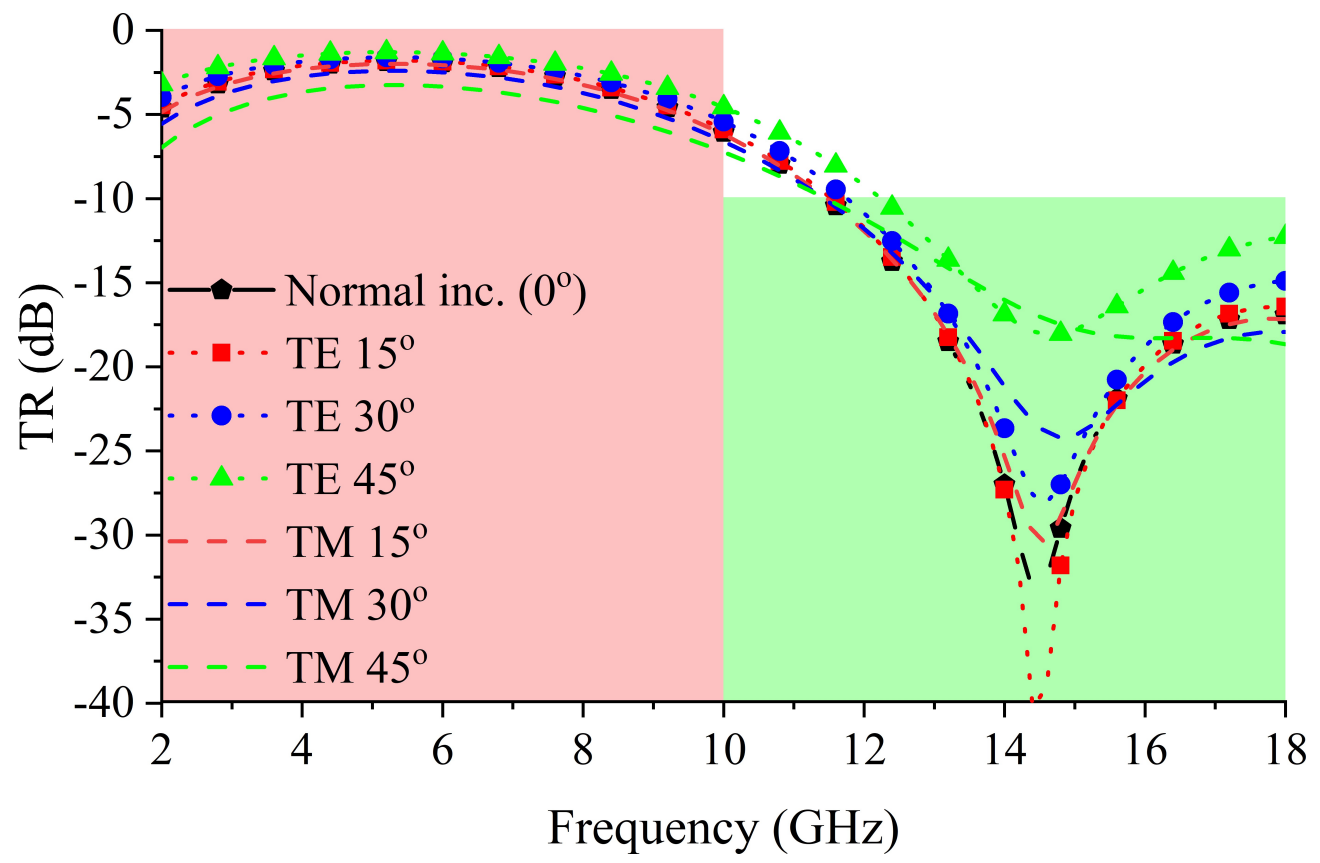

(b)

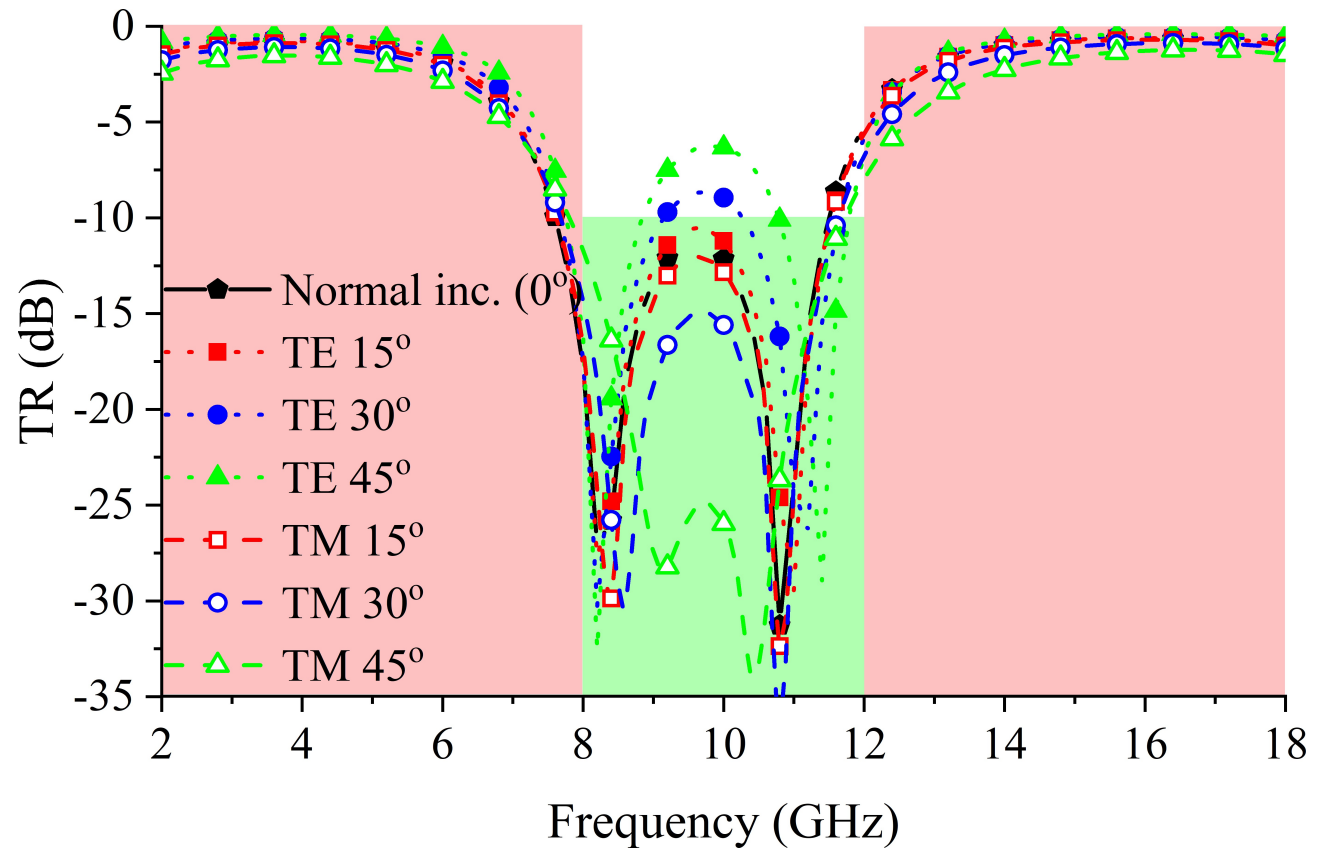

(c)

**Fig. 16.4.** The TR plots of the global optimal solution for the design of BP filter

In order to elaborately examine the TR results of the MMDFs, the maximum, average and minimum values of the TRs at the pass and stop band regions are obtained for the TE and TM polarizations at 0, 15, 30 and 45 degrees in Tables 16.3, 16.4 and 16.5. The performance of the multi-objective optimization scheme based on ABC is even clearly seen from the results in the tables. It is also verified that the TRs for TE are always higher than the TM. The average TRs are generally lower than -10 dB at the pass bands and higher than -3 dB at stop bands at which level 50% of the incident wave power reflects from the MMDF.

**Table 16.3.** The maximum, average and minimum TR of the LP filter at the pass and stop bands

| Band | Polarization-incidence | | TR | | |
|---|---|---|---|---|---|
| | | | Maximum | Average | Minimum |
| Pass | 0° | | -4.47 | -14.86 | -25.64 |
| | TE | 15° | -4.60 | -14.35 | -23.77 |
| | | 30° | -4.97 | -12.83 | -20.87 |
| | | 45° | -5.39 | -10.38 | -16.92 |
| | TM | 15° | -5.13 | -15.38 | -26.99 |
| | | 30° | -6.93 | -17.14 | -34.17 |
| | | 45° | -8.64 | -18.76 | -31.40 |
| Stop | 0° | | -2.18 | -2.36 | -3.52 |
| | TE | 15° | -2.14 | -2.32 | -3.60 |
| | | 30° | -1.99 | -2.21 | -3.80 |
| | | 45° | -1.70 | -1.98 | -3.97 |
| | TM | 15° | -2.29 | -2.51 | -4.05 |
| | | 30° | -2.62 | -3.04 | -5.62 |
| | | 45° | -3.21 | -4.00 | -7.46 |

**Table 16.4.** The maximum, average and minimum TR of the HP filter at the pass and stop bands

| Band | Polarization-incidence | | TR | | |
|---|---|---|---|---|---|
| | | | Maximum | Average | Minimum |
| Stop | 0° | | -1.87 | -2.94 | -5.66 |
| | TE | 15° | -1.80 | -2.84 | -5.51 |
| | | 30° | -1.61 | -2.56 | -5.05 |
| | | 45° | -1.30 | -2.10 | -4.25 |
| | TM | 15° | -1.99 | -3.08 | -5.80 |
| | | 30° | -2.41 | -3.56 | -6.22 |
| | | 45° | -3.25 | -4.48 | -6.96 |
| Pass | 0° | | -6.49 | -17.89 | -33.35 |
| | TE | 15° | -6.33 | -18.19 | -40.28 |
| | | 30° | -5.81 | -16.20 | -28.09 |
| | | 45° | -4.91 | -12.66 | -18.09 |
| | TM | 15° | -6.62 | -17.66 | -30.31 |
| | | 30° | -6.99 | -16.82 | -24.21 |
| | | 45° | -7.58 | -14.80 | -18.66 |

**Table 16.5.** The maximum, average and minimum TR of the BP filter at the pass and stop bands

| Band | Polarization-incidence | | TR | | |
|---|---|---|---|---|---|
| | | | Maximum | Average | Minimum |
| Stop | 0° | | -0.65 | -1.87 | -13.04 |
| | TE | 15° | -0.62 | -1.81 | -12.82 |
| | | 30° | -0.54 | -1.62 | -12.10 |
| | | 45° | -0.41 | -1.34 | -10.71 |
| | TM | 15° | -0.70 | -1.95 | -12.57 |
| | | 30° | -0.87 | -2.21 | -11.41 |
| | | 45° | -1.23 | -2.70 | -9.95 |
| Pass | 0° | | -6.63 | -16.04 | -31.27 |
| | TE | 15° | -6.89 | -15.55 | -29.46 |
| | | 30° | -7.79 | -14.27 | -30.30 |
| | | 45° | -6.24 | -12.43 | -32.56 |
| | TM | 15° | -7.12 | -16.86 | -32.37 |
| | | 30° | -8.33 | -19.14 | -36.90 |
| | | 45° | -9.37 | -21.94 | -34.00 |

It can be inferred from results given in Fig. 16.4, Tables 16.3, 16.4 and 16.5 that ABC is even successful in multi-objective EM problems. Since ABC is one of the latest metaheuristic algorithms, its implementations to multi-objective engineering design problems are not as much as the relatively former algorithms such as GA, PSO and DE. It is observed that multi-objective variants of ABC have been attempted to developed as testing for benchmark functions [30, 31] and the results are better than those of the former algorithms. Therefore, the multi-objective versions of ABC should be implemented to more engineering problems in order to demonstrate its superiority over the former algorithms.

## 1.4 Conclusions

Single objective ABC is a simple and robust meta-heuristic optimization algorithm based on the collective foraging behavior of the natural honey bees. On the other hand, the multi-objective performance of ABC on engineering design problems especially on computational electromagnetic needs more implementations and demonstrations. Efficacy on the design of MMDF including expensive computations is anticipated through a multi-objective scheme. In this chapter, three types of MMDF which are LP, HP and BP filters are designed through the MO-ABC scheme. Two objective functions involving mean TR at the pass band and stop band are constituted across an exact EM model based on the reflection mechanism taking into account even the incident wave angle with polarizations TE and TM. The Pareto optimal solutions are refined within the possible solutions for synchronously minimizing the objective functions. The global optimal MMDF designs are selected from the Pareto optimal solutions by providing the trade-off between the objective functions. The performance results of the designed MMDFs are investigated with regard to the frequency characteristics. Therefore, the designed MMDFs show near ideal filter characteristics thank to MO-ABC.

***Acknowledgement.*** The author is thankful to Dr. Deniz Ustun for valuable support and contribution to the computations carried out in this study.